\documentclass[twocolumn,prb,showpacs,superscriptaddress]{revtex4}
\usepackage{epsfig}
\begin{document}
\topmargin-1.0cm

\title {
Hydrostatic pressure dependence of the luminescence and Raman
frequencies in polyfluorene}
\author {C.M. Martin}\affiliation {Department of Physics,
University of Missouri, Columbia, MO 65211 USA}
\author{S. Guha}\email[Corresponding author E-mail:]{sug100f@smsu.edu}
\affiliation {Department of Physics, Astronomy and Materials
Science, Southwest Missouri State University, Springfield, MO
65804 USA}
\author {M. Chandrasekhar} \author{H.R. Chandrasekhar}
\affiliation {Department of Physics, University of Missouri,
Columbia, MO 65211 USA} \author{ R. Guentner} \author{P.
Scanduicci de Freitas} \author{ U. Scherf} \affiliation {Institut
f\"ur Chemie and Polymerchemie, Universit\"at Potsdam, Germany}

\date{\today}

\begin{abstract}
We present studies of the photoluminescence (PL), absorption and
Raman scattering from poly[2,7-(9,9'-bis(2-ethylhexyl))fluorene]
under hydrostatic pressures of 0-100 kbar at room temperature. The
well-defined PL and associated vibronics that are observed at
atmospheric pressure change dramatically around 20 kbar in the
bulk sample and at around 35 kbar for the thin film sample. Beyond
these pressures the PL emission from the backbone is swamped by
strong peaks due to aggregates and keto defects in the 2.1-2.6 eV
region. The Raman peaks shift to higher energies and exhibit
unexpected antiresonance lineshapes at higher pressures,
indicating a strong electron-phonon interaction.

\end{abstract}

 \maketitle

\section{Introduction} \label{sec:intro}
Polyfluorenes (PF) have emerged as attractive alternatives\cite
{leclerc, scherf} to other conjugated polymers for display
applications due to their efficient blue emission and high hole
mobility at room temperature.\cite{grice, redecker} Blue
electroluminescent materials are of particular interest for
organic displays since blue light can be converted into red and
green quite easily by color-changing media (fluorescent
dyes).\cite{tasch} PFs have been very efficiently utilized in
organic light emitting diodes (OLEDs).\cite{bernius} They have
been found to exhibit complex morphological behavior that has
interesting implications due to their rich photophysics. The
liquid crystallinity of PF derivatives with long alkyl
substituents allows fabrication of electroluminescent devices with
highly polarized emission\cite{miteva}  that is of potential use
for backlighting in liquid crystalline displays.\cite{grell}

OLEDs fabricated from the PF family of materials are known to
degrade under operation.  The desired blue electroluminescence
from the singlet excitons changes to an undesirable 2.1-2.6 eV
greenish/reddish emission. Photo-oxidation produces a similar
broad emission band in the photoluminescence (PL) spectrum. This
red shifted emission has been attributed to aggregation and/or
excimer formation in the material.\cite{carter,fujikawa} Recently,
List \textit{et al}.\cite{list} have conclusively shown that the
2.3 eV band is related to emission from  keto defect sites
(9-fluorenone) in the sample. These defect sites act as guest
emitters that can efficiently trap singlet excitons created on the
conjugated polyfluorene backbone. Time-resolved PL measurements
further show different temporal dynamics for the various peaks in
the 2.1-2.6 eV region, indicating that the origin of some of the
peaks is due to on-chain emissive defects while other peaks are
from aggregates and excimers.\cite{lupton,herz}

In this work we probe the optical properties of
poly[2,7-(9,9'-bis(2-ethylhexyl))fluorene] (PF2/6) via PL,
absorption, and Raman scattering as a function of hydrostatic
pressure. Hydrostatic pressure enhances intermolecular interaction
and changes the molecular geometry without producing chemical
changes. An understanding of the influence of the intermolecular
interactions is crucial since the photoluminescence quantum yield
(PLQY) of polymers is known to decrease in the solid state. For
example, in methylated ladder-type poly {\em para}-phenylene
(m-LPPP), the PLQY of solution and film are (100\%) and (30\%),
respectively.\cite{tasch2} In decyloxy PPP (DOPPP) PLQY values of
85\% in solution and 35\% in film are observed.\cite{heeger} In
the solid state the electronic properties of organic materials
depend significantly on the three-dimensional interactions. Recent
theoretical methods of quantum-chemistry and solid-state physics
in conjunction with experimental measurements have provided
valuable insight into the electronic and optical properties of
both isolated and interacting conjugated chains (oligomers and
polymers).\cite{beljonne,cornil,spano,puschnig} Hydrostatic
pressure studies also allow us to probe the effects of enhanced
interaction on the aggregate and defect-related emissions in these
systems.

An additional interesting feature of PF2/6 is that its backbone
conformation is intermediate to that of planar and non-planar
conjugated solids. It thus provides an interesting contrast to the
planar polymer m-LPPP and the non-planar oligophenyls such as {\em
para} hexaphenyl (PHP). Our previous studies indicate that
enhanced intermolecular interactions under the impact of pressure
in conjugated solids typically produce an increased degree of
conjugation, destabilization of localized states as in m-LPPP,
\cite{yang, meerapressure}  and changes in the ring torsional
motion as in PHP.\cite{guha} These changes in geometry can be
deduced from their influence on electronic and vibrational
spectra. PFs also show rich morphological behavior as a function
of temperature: the glass-rubber transition and the liquid
crystalline phase are induced at elevated temperatures. It
therefore makes it an interesting system to study under high
pressure since one can test whether any of the high-temperature
phases can be induced by the application of pressure alone at room
temperature, thus allowing the study of the corresponding changes
in their electronic and vibrational properties in a different
region of phase space.

\section{Experimental Details}\label{sec:exptaldetails}

\subsection{Methodology}\label{sec:expdetailsmethod}
Pressure studies were conducted in a Merrill-Bassett type diamond
anvil cell (DAC) with cryogenically loaded argon as the pressure
medium.  A small chip of ruby was included in the sample chamber,
thus allowing the determination of pressure by measuring the shift
of the luminescence spectrum of the ruby.  For ambient pressure
measurements the samples were loaded in a cryostat that was
evacuated to below 100 mTorr to prevent any photo-oxidative
damage.

 PL and absorption spectra were recorded using an Ocean Optics PC2000
spectrometer with 25-$\mu$m slits. The 351.1nm  line from an $\rm
Ar^+$ laser was used in a backscattering configuration for PL.
Raman measurements were performed using a Spex triplemate
spectrometer and the 647.1 nm line of a $\rm Kr^+$ laser with 10
mW incident power in a backscattering configuration.  The
scattered light was detected using a cryogenically cooled CCD
array detector and a holographic super-notch filter to block the
elastically scattered light. Measurement of the spectrum of a chip
of ruby outside the pressure chamber before each Raman measurement
ensured that the calibration of the spectrometer was consistent
from one day to the next. The low temperature measurements were
conducted in a closed cycle helium refrigerator. The spectral data
was analyzed using PeakFit and Origin to determine the position,
area, and full width at half maximum (FWHM) of the peaks.

\subsection{Sample Details}\label{sec:expdetailspf2/6}

PF2/6 forms planar monomer units but has a torsional degree of
freedom between adjacent monomer units as shown in the inset of
Figure 1 (b); its synthesis is described in Ref.
(\onlinecite{scherf}). The torsional freedom in PF2/6 makes the
structure an intermediate case between non-planar oligophenyls and
planar m-LPPP. PF2/6 is semi-crystalline at room temperature, with
a glass-rubber transition temperature  (T$_g$) of the amorphous
component at 80 $\rm ^o$C, and a crystalline to nematic liquid
crystalline phase transition at 167 $\rm ^o$C (reverse transition
at 132 $\rm ^o$C upon cooling).\cite{scherf} The molecules have
been shown to have a C5 helical conformation at room temperature.
\cite{scherf} Upon further processing some polyfluorene films
display a $\beta$ phase, which has a more extended intrachain
$\pi$-conjugation, in addition to the regular glassy $\alpha$
phase. The $\beta$ phase has been detected in 9,9-di-n-octyl-PF
(PF8 or PFO)\cite{grell,cadby} and shows a distinct red shift of
absorption and emission peaks with a well-resolved vibronic
progression both in absorption and emission. In contrast, the
$\alpha$ phase shows a well-resolved vibronic progression only in
the emission spectrum. Due to the branched alkyl side groups this
material is not expected to easily form a $\beta$ phase, however,
it has been shown that a highly aligned state can be achieved with
solvent treatment and thermal cycling .\cite{lieser}

The PL and Raman spectra were measured from a powder sample of
PF2/6, which has both amorphous and crystalline components, while
the absorption and PL were measured from a film. The film was
prepared by drop-casting PF2/6 dissolved in spectroscopic grade
dichloromethane directly onto the surface of the bottom diamond of
the DAC. While the film is loosely attached to the diamond
surface, it is not bonded onto it, and therefore experiences
hydrostatic pressure in the DAC.

All the Raman measurements were carried out with a bulk (powder)
sample. This was necessary because the Raman signal from a film is
very weak due to the drastically reduced scattering volume.

\section{Steady-State Photoluminescence and Absorption}\label{sec:PL}


\subsection{Experimental Results}\label{sec:PLresults}
\begin{figure}
\unitlength1cm
\begin{picture}(5.0,6.2)
\put(-2.,-0.2){ \epsfig{file=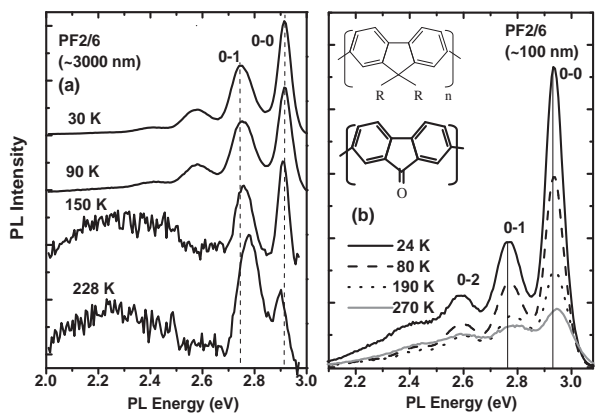, angle=0, width=9.2cm,
totalheight=7.3cm}}
\end{picture}
\caption{PL spectrum of PF2/6 at selected values of temperature
for (a) a thick and  (b) a thin film. The vertical lines indicate
the shift in the transition energies as temperature is increased.
The top inset in (b) shows a monomer structural unit of PF2/6
where R denotes the ethyl hexyl side group and the bottom inset is
the fluorenone structural unit (keto defect).} \label{figure1}
\end{figure}

Figure \ref{figure1} shows the PL spectra from two PF2/6 films
(panel (a) thick film, $\approx$3000nm; panel (b) thin film,
$\approx$100nm) for a few selected values of temperature. Vibronic
progressions are seen in the PL emission of both films, indicating
a coupling of the backbone carbon-carbon stretch vibration to the
electronic transitions. The vibronic peaks result from a non-zero
overlap of different vibronic wavefunctions of the electronic
ground and excited states. The transition highest in energy is the
0-0 transition, which takes place between the zeroth vibronic
level in the excited state and the zeroth vibronic level in the
ground state. The 0-1 transition involves the creation of one
phonon. The relative intensity of the 0-0 peak to the 0-1 peak in
the thick film (Figure \ref{figure1}(a)) is lower than that in the
thin film (Figure \ref{figure1}(b)) indicating a higher
self-absorption in the thick film. The peak positions were
determined by fitting the spectra to Gaussian lineshapes. In the
thin film, the main vibronic peaks that are observed at 30 K are
the 0-0 peak at 2.93 eV, the 0-1 at 2.77 eV and the 0-2 transition
at 2.59 eV. An additional vibronic replica is observed at 2.86 eV
between the 0-1 and 0-2 peaks. These values are very close to the
energies observed in the thick film.

With increasing temperatures a broad peak is seen to emerge at 2.3
eV around 150 K for the thick film.  Recent work suggests that
this peak is related to emission from keto defect sites,
\cite{list, scherf} which are due to the inclusion of fluorenone
units along the PF backbone. Keto defects (shown in the inset of
Figure \ref{figure1}(b)) can be accidentally incorporated into the
$\pi$-conjugated PF backbone either during synthesis, by direct
inclusion of a fluorenone unit, due to oxidization of
non-alkylated or monosubstituted fluorene sites, or as a result of
a photo-oxidative degradation process. The concentration of these
defect sites is quite low in our PF2/6 sample since the 2.3 eV
emission is absent in the thin film (Figure \ref{figure1}(b)).
Also, it is a thermally activated process; a weak defect-related
emission is only observed for temperatures above 150 K for the
thick film (Figure \ref{figure1}(a)).

\begin{figure}
\unitlength1cm
\begin{picture}(4.8,6.2)
\put(-3.3,-0.5){ \epsfig{file=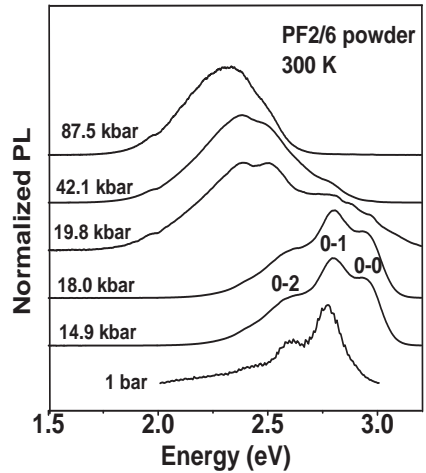, angle=0, width=7.5cm,
totalheight=9.cm}}
\end{picture}
\caption{PL spectrum of bulk PF2/6 at selected values of
hydrostatic pressure. The 1 bar spectrum was measured from a
sample evacuated to below 100 mTorr to prevent photo-oxidative
damage.} \label{figure2}
\end{figure}

Figure \ref{figure2} shows the 300 K PL spectrum of a bulk powder
sample of PF2/6 at selected values of pressure. The 0-0 transition
(2.9 eV) at ambient pressure is barely visible due to
self-absorption effects. However, the other vibronics at ambient
pressure correlate with the film data in Figure \ref{figure1}. The
most remarkable feature is the emergence of a strong orange
emission above $\sim$20 kbar at 2.4 eV, which dominates the
spectrum and completely overwhelms the PL from the backbone. This
peak is attributed to a combination of emissions from aggregates
and keto defect sites (discussed in greater detail in Section
\ref{sec:PLdiscussion}). The broad emission, which is comprised of
at least 3 peaks, clearly red shifts with increasing pressures.
The pressure dependence of the backbone PL transitions is
difficult to quantify. The apparent position of the 0-0 transition
is affected by changes in the absorption spectrum, and the
relative intensities of the backbone and aggregate/defect
emissions change dramatically as pressure is increased. The
vibronic peaks are clearly observable up to 20 kbar above which
pressure they appear to remain relatively constant in energy with
increasing pressure. This behavior is contrary to the observations
in other $\pi$-conjugated molecules and polymers where the
backbone emission clearly red shifts under increasing pressures
as, for example, in polyacetylene,\cite{moses}
polythiophene,\cite{hess} m-LPPP,\cite{yang} PHP,\cite{guha2} poly
(p-phenylene vinylene) (PPV)\cite{webster} and
MEH-PPV.\cite{drickamer}

\begin{figure}
\unitlength1cm
\begin{picture}(5.0,6.)
\put(-2.,-0.5){ \epsfig{file=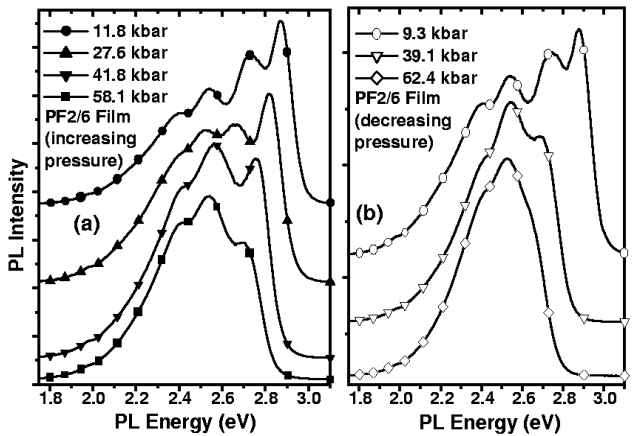, angle=0, width=9.7cm,
totalheight=7.3cm}}
\end{picture}
\caption{Room temperature PL spectrum of a PF2/6 film at selected
values of pressure (a) while increasing the pressure and (b)on
lowering the pressure.} \label{figure3}
\end{figure}

In order to clarify the behavior of the backbone emission which,
as seen from Figure \ref{figure1}, is less affected by the keto
defect in thin films, we studied the pressure dependence of the PL
from a film of PF2/6 that was drop-cast onto the surface of one of
the diamonds of the DAC. We systematically measured the PL both
for increasing and decreasing pressures. The backbone emission is
clearly defined up to 30 kbar, as shown in Figure \ref{figure3}
(a). Above 30 kbar the 0-1 and 0-2 can no longer be clearly
distinguished; instead a broad peak appears at 2.5 eV with a
shoulder at 2.4 eV. The 0-0 vibronic peak is clearly observed up
to 42 kbar as shown in Figure \ref{figure3} (a). As the pressure
increases further the 2.5 eV peak dominates the spectrum up to the
highest pressures ($\sim$100 kbar). Figure \ref{figure3} (b) shows
the PL spectra taken while lowering the pressure. Some hysteresis
is observed in recovering the low-pressure spectra. In the
mid-range (30-45 kbar) the emission in the 2.2-2.5 eV is still
dominant and the backbone emission recovers very slowly. Full
recovery is not observed until $\sim$10 kbar. We note that our
results for the PF2/6 film under hydrostatic pressure conditions
in the DAC are quite different from that observed by Rothe
\textit{et al}.\cite{rothe}, where they observe no hysteresis. In
their work a film of PF2/6 was sandwiched between two glass
slides: the pressure in their case should be substantially lower
than that in a DAC, and is uniaxial as well.

\begin{figure}
\unitlength1cm
\begin{picture}(5.0,6.)
\put(-2.4,-0.5){ \epsfig{file=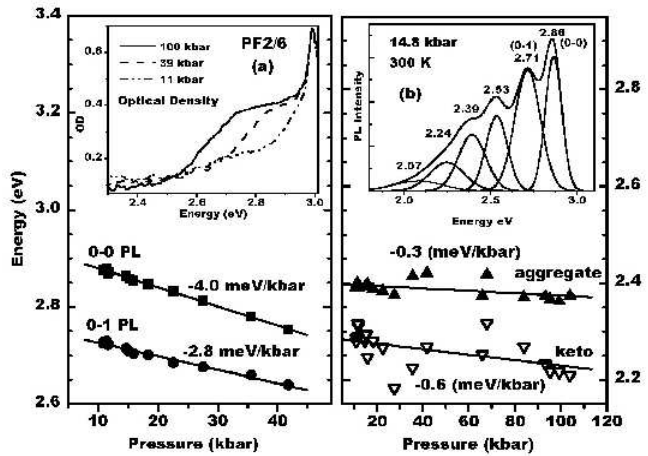, angle=0, width=8.9cm,
totalheight=6.5cm}}
\end{picture}
\caption{Peak positions of the various transitions in a film of
PF2/6 as a function of increasing pressure at 300 K. The peak
position of the the 0-0 and the 0-1 backbone PL as a function of
pressure are shown in (a) and the aggregate and the keto peak
positions are shown in (b). The slopes of the various transitions
are indicated in the figure. The inset in panel (a) shows the
absorption spectra at selected values of pressure. The inset in
panel (b) shows the lineshape fits for the 14.8 kbar PL spectrum.
The peak positions are in eV.} \label{figure4}
\end{figure}

The absorption spectrum as a function of pressure was measured
from the same film (at the same pressures as the PL) and is shown
in the inset of Figure \ref{figure4}. The spectra were taken at
room temperature by dividing the sample transmission by the
transmission of the empty diamond cell. Such a broad absorption
spectrum is characteristic of the $\alpha$ phase. With increasing
pressure, the overall absorbance increases and the spectrum red
shifts. Increased absorbance in the 2.7 eV region may be
indicative of $\beta$ phase formation. Although the $\alpha$ phase
of PF does not show any absorption below 2.90 eV, one cannot rule
out a red shift of the absorption edge from the $\alpha$ phase
itself with increasing pressures. Such effects have been observed
in m-LPPP under pressure. \cite{meerapressure} The sharp peak at
2.95 eV is from the type II diamond of the pressure cell.

\subsection{Discussion of PL Results}\label{sec:PLdiscussion}
The PL spectrum of the film, was fit using six Gaussian peaks. In
the insert of Figure 4 (b) we show a spectrum at 14.8 kbar, where
the vibronics, keto, and aggregate emissions are all clearly
observed. The highest two are the 0-0 and 0-1 vibronics. The 2.53
eV peak is close to the 0-2 position. The 2.39 eV peak arises from
an aggregate, as does the 2.07 peak. The 2.24 eV peak arises from
the keto defect. The peak positions of the keto and the aggregate
emissions versus pressure for the film sample are plotted in
Figure \ref{figure4} (b), and they red shift by 0.6 and 0.3
meV/kbar, respectively, with increasing pressure. The 0-0 and the
0-1 backbone vibronics are clearly visible up to 40 kbar and shift
by about an order of magnitude faster (Figure \ref{figure4} (a)).
Similar systematic fitting of the backbone vibronic peaks is
difficult for the powder sample, due to the increasing absorbance
and the broad keto emission at higher pressures.

The main impact of pressure on the PL spectrum is the enhanced
emission in the 2.1-2.6 eV range which overwhelms the backbone
emission of PF2/6 beyond a certain pressure value. Emission in
this region has been observed in other works on PF at ambient
pressure. Under photodegradation in
9,9-di-n-hexyl-2,7-dibromofluorene (PDHF), three peaks at 2.2 eV,
2.4 eV and 2.6 eV are observed and have been identified as excimer
peaks.\cite{carter} Using time-resolved PL spectroscopy, Herz and
Phillips\cite{herz} find a longer decay time for the 2.15 eV peak,
also attributed to an excimer type emission, as compared with
other peaks in this region. This result has led to a picture that
intermolecular interactions dominate the bulk material properties
in PFs.

Work by List \textit {et al.}\cite{list} conclusively shows that
the 2.2-2.3 eV emission is from keto defect sites in PFs. This
assignment has been confirmed in a recent work by Lupton \textit
{et al.}\cite{lupton} using time-resolved PL on a range of
poly-and oligofluorene samples. They attribute the 2.3 eV emission
to an on-chain emissive defect, in agreement with the keto-type
defect. Further, they observe an enhancement of the 2.6 eV
emission with increasing sample concentration concluding that the
origin of this peak is due to the formation of interchain
aggregates.

In our work here, the enhancement of the 2.1-2.6 eV emission for
the bulk sample occurs at $\sim$ 20 kbar and for the film at a
slightly higher pressure of $\sim$ 35 kbar. The relative
intensities of the various peaks in the 2.1-2.6 eV band under
pressure are different in the film and in a powder sample. In bulk
PF2/6 (see Figure \ref{figure2}) both the keto and aggregate
emission are equally enhanced at 20 kbar: the keto peak ($\sim$
2.3 eV) gains intensity over the aggregate peak ($\sim$ 2.4 eV)
beyond 42 kbar. Although the concentration of the keto defects is
quite small in our PF2/6 sample (as shown in Figure
\ref{figure1}), increasing the intermolecular interaction causes
the emission from the defect sites to be considerably enhanced.
This scenario is somewhat similar to the guest-host polyfluorene
system where a small concentration of a chromophore with an
absorption spectrum closely matching the emission of PFO
drastically changes the PL emission.\cite{virgili,cerullo} In
these systems there is an efficient F\"{o}rster energy transfer to
the host molecules that allows conversion of the blue emission
from PFO into red emission of the guest chromophore. In our case
the enhanced intermolecular interaction allows transfer from the
backbone to the keto defect making that emission more and more
prominent as pressure increases.

In the PF2/6 film, beyond 35 kbar the 2.1-2.6 eV emission
dominates the backbone emission. The aggregate emission is
stronger than the keto emission which appears as a weak shoulder
at higher pressures. The higher aggregation in the film may result
from a more compact and ordered morphology in the film. This
ordering may also explain the hysteresis of the PL spectra
observed between increasing and decreasing pressures. The absence
or weak intensity of keto-related peaks in the film may result
from a ''pinning'' of excitations in ordered regions that prevents
diffusion of the excited carriers to the defect states.

A weak shoulder at 1.9 eV is observed both in the film and powder
samples of PF2/6 (see Figures \ref{figure2} and \ref{figure3}).
The exact origin of this peak is not understood but most probably
it is also due to aggregate emission. In Section
\ref{sec:ramanscatter} we discuss how this PL emission at higher
pressures results in an antiresonance effect of the Raman phonons.

The overall PL intensity, due to aggregate and keto emission,
increases noticeably at 42 kbar in the powder sample. A
temperature scan was taken at this pressure since cooling the
sample decreases the pressure by $\sim$ 2 kbar, and thus allows us
to cycle through any morphological changes. The relative intensity
of the backbone transitions was observed to decrease with
increasing temperature. A significant change occurs between 150 K
and 250 K. Above 250 K the 0-0 transition is almost completely
quenched and the broadband emission increases in intensity by more
than a factor of three. These changes in the PL are most likely
due to morphological changes in PF2/6.\cite {michael} Transport
measurements show that polymers in a liquid crystalline state
exhibit greatly enhanced interchain charge carrier
mobility.\cite{redecker2} Since PF2/6 does exhibit a crystalline
to nematic liquid crystalline phase transition, it is possible
that such a transition causes the sudden change in the relative
intensities of the PL emission with increasing pressures.

The rate of red-shift of the 0-0 and 0-1 backbone emission peaks
in the PF2/6 film under pressure is similar to shifts observed in
m-LPPP film and PHP powder, as shown in Table \ref{table1}. Future
work on copolymers of PF2/6 and with other side-group
substitutions (that minimize the incorporation of keto defects)
are required to understand the dynamics of the backbone vibronics
in the bulk PF2/6 sample, which barely appear to shift. This
effect may well be an artifact due to the higher self-absorption
in the powder or due to a different morphology of the powder as
compared with the film under hydrostatic pressure. The shifts of
the aggregate and keto emissions with pressure are similar in film
and powder samples.

\begin{table}
\caption{Pressure coefficients for backbone PL emission peaks in
PF2/6 film, m-LPPP film, and PHP powder (beyond 15 kbar) for the
0-0 and 0-1 PL peaks. The pressure coefficients are determined by
a linear fit to the PL energy positions vs pressure given by
$E(P)=E(0)+\alpha P$.}

\label{table1}
\begin{ruledtabular}
\begin{tabular}{cccc}
Sample        & $\alpha_{0-0}$(meV/kbar) &
  $\alpha_{0-1}$ (meV/kbar)\\
\hline PF2/6 Film\footnotemark[1]     & -4.0 $\pm$ 0.1 & -2.8$\pm$ 0.1\\
m-LPPP film\footnotemark[2] & -2.5 $\pm$ 0.1 & -2.1
$\pm$ 0.1\\
PHP powder \footnotemark[2] & -4.4 $\pm$ 0.2        & -5.3 $\pm$ 0.1\\
\footnotetext[1]{this work}
\footnotetext[2]{Ref.\onlinecite{guha2}}

\end{tabular}
\end{ruledtabular}
\end{table}


\section{Raman scattering}\label{sec:ramanscatter}

\subsection{Experimental Results}

\begin{figure}
\unitlength1cm
\begin{picture}(5.0,7.0)
\put(-2.,0.4){ \epsfig{file=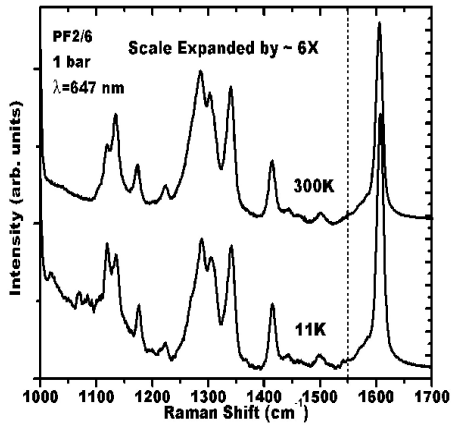, angle=0, width=7.2cm,
totalheight=6.8cm}}
\end{picture}
\caption{Raman spectra of PF2/6 at ambient pressure at 13 K and
300 K. The vertical scale has been expanded by about 6X below
1550cm$^{-1}$. } \label{figure5}
\end{figure}

Figure \ref{figure5} shows the Raman spectra of powder PF2/6 at 13
K and 300 K at ambient pressure. With increasing temperatures all
the Raman peaks in the 1200-1600 cm$^{-1}$ soften by  $\sim$ 2
cm$^{-1}$, a value comparable to several known conjugated
materials.\cite{guha}

The Raman frequencies in the 1050-1200 cm$^{-1}$ region are
sensitive to side group substitution; they arise from the C-H
bending modes of the ethyl-hexyl side group in PF2/6. The 1290
cm$^{-1}$, 1342 cm$^{-1}$ and the 1417 cm$^{-1}$ are associated
with the backbone C-C stretch modes. The Raman peaks in the 1600
cm$^{-1}$ region arise from the intra-ring C-C stretch mode. This
region of the spectrum is best fit with two peaks, a weak peak at
1582 cm$^{-1}$ and a strong peak at 1605 cm$^{-1}$. These peak
assignments are similar to those of Ariu \textit {et
al.}\cite{ariu} in their work on PFO, where the Raman peaks in the
1200-1400 cm$^{-1}$ region are attributed to the C-C stretch mode
between adjacent phenyl rings. It has been assumed that the higher
frequency modes in this region are due to the C-C stretch from
within the more rigid monomer unit, while the C-C stretch modes
between adjacent monomer units are expected to have a lower
frequency due to the torsional degree of freedom
allowed.\cite{ariu}

\begin{figure}
\unitlength1cm
\begin{picture}(5.0,6.)
\put(-2.,-0.5){ \epsfig{file=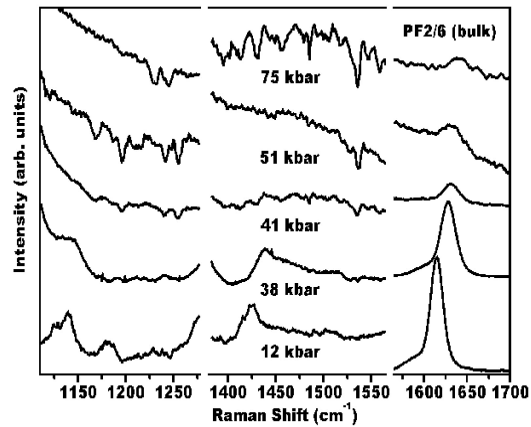, angle=0, width=7.2cm,
totalheight=6.5cm}}
\end{picture}
\caption{Raman spectra of PF2/6 at selected pressures. Each
frequency region of the graph has been scaled individually to
highlight the key features. Owing to the rising PL with increasing
pressure, the background increases at higher pressures obscuring
the the low frequency Raman peaks.} \label{figure6}
\end{figure}

Figure \ref{figure6} is a plot of Raman spectra from PF2/6 at
various pressures. Each frequency region of the graph has been
scaled individually so that key features in each spectrum are
clearly observed. The break in the $\it{x}$-axis at ~1300
cm$^{-1}$ denotes the region where the Raman peak from the diamond
has been removed for clarity. Most of the phonon frequencies
corresponding to the backbone C-C stretch modes are not observable
as a function of pressure due to the presence of the strong ~1330
cm$^{-1}$ Raman peak from diamonds in the DAC.

As pressure is increased, all of the Raman frequencies analyzed
tend to increase in frequency. Concurrently, the PL tail gains
intensity, obscuring the Raman signal beginning with the lower
frequency peaks. The region below $\sim$ 1300 cm$^{-1}$ is
observable up to $\sim$ 20 kbar and the 1417 cm$^{-1}$ peak is
observed up to about 40 kbar. We note that these Raman
measurements were taken using the 647.1 nm (1.92 eV) line of a
Kr$^{+}$ laser. Despite being well below the PL peaks of both the
backbone and the keto emissions, the laser line still excites the
low energy tail of the PL spectrum (see Figure \ref{figure3}). We
believe that this tail most likely arises from aggregate emissions
but possibly could also be due to the keto defects.

The rising PL tail above 20 kbar causes several of the Raman peaks
to exhibit asymmetric lineshapes and an antiresonance effect
characteristic of a Breit-Wigner Fano (BWF) resonance. Such
effects have been observed in inorganic semiconductors, for
example in p- and n-doped Si and Ge.\cite{meera,cardona} This
effect is indicative of a strong electron-phonon interaction
between the Raman phonons and the electronic
continuum.\cite{klein} We find that the most consistent
development of the BWF resonance is observed in the 1605 cm$^{-1}$
peak with the development of asymmetry and increased broadening
with increasing pressures is seen in Figure 6. We have conducted a
detailed lineshape analysis for this peak in terms of the BWF
resonance, discussed in detail in section
\ref{sec:ramandiscussion}.

\begin{figure}
\unitlength1cm
\begin{picture}(6.2,7.2)
\put(-1.2,0.){ \epsfig{file=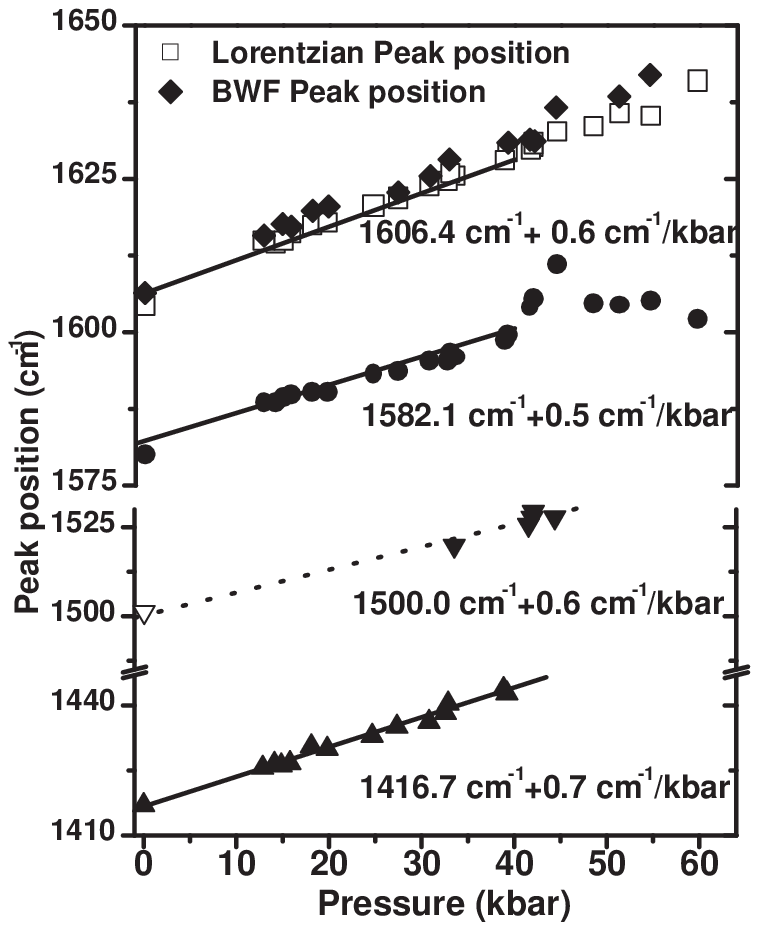, angle=0, width=6.6cm,
totalheight=8.0cm}}
\end{picture}
\caption{Peak positions of the 1417 cm$^{-1}$, 1500 cm$^{-1}$
(IR-active), 1582 cm$^{-1}$ (shoulder of 1605 cm$^{-1}$), and the
1605 cm$^{-1}$ modes as a function of pressure. The open symbol
($\nabla$) denotes the position of the 1500 cm$^{-1}$ peak at 1
bar. The slopes for the linear fits have an uncertainty of $\pm$
0.05 cm$^{-1}$/kbar. The straight lines for the 1605 cm$^{-1}$ and
the 1582 cm$^{-1}$ peaks are fits to the frequencies obtained from
the Lorentzian fits up to 40 kbar.} \label{figure7}
\end{figure}

Along with the asymmetry of the 1605 cm$^{-1}$ peak, new peaks
exhibiting antiresonance effects appear near ~1230 cm$^{-1}$ and
1530 cm$^{-1}$ (as seen in Figure 6). These peaks appear in the
Raman spectra both with increasing and decreasing pressure.
Although their signals are weak, their spectra are reproducible.
Some of them may be infrared (IR) active modes that become Raman
active at higher pressures. In Ref. (\onlinecite {list}) it is
seen that there are a number of IR peaks in the 1000-1500
cm$^{-1}$ range in PF with the strongest IR peak appearing at
~1500 cm$^{-1}$. The 1530 cm$^{-1}$ feature observed around 40
kbar is most likely the 1500 cm$^{-1}$ IR active peak shifted to
higher frequencies due to the applied pressure. This IR frequency
shows a weak signature even in the 1 bar Raman spectrum (Figures
\ref{figure5} and \ref{figure6}), and the magnitude of the shift
is consistent with the pressure shifts of the other Raman peaks,
as seen in Figure \ref{figure7}.

%
%
\subsection{Discussion of Raman Results}\label{sec:ramandiscussion}
The Raman peaks have been fit with Lorentzian lineshapes to
determine their frequency positions. Their peak positions as a
function of pressure are plotted in Figure \ref{figure7}. The 1417
cm$^{-1}$ mode is seen clearly up to 40 kbar. The 1500 cm$^{-1}$
peak is too weak in intensity to track below 40 kbar, but can be
seen as a marked antiresonance and can be tracked reliably for a
few data points above 40 kbar. The 34 kbar data point for this
peak was taken as the pressure was being decreased, and due to a
hysteresis effect the peak was still visible. As the pressure was
decreased further the PL background dropped in intensity and the
low pressure spectra were fully recovered. All of the peaks shift
with fairly similar pressure coefficients.

The positions of the 1605 cm$^{-1}$ and the shoulder peak at 1582
cm$^{-1}$ are included up to 60 kbar in Figure \ref{figure7}.
Above about 40 kbar the increase of the asymmetry of the 1605
cm$^{-1}$ peak made fitting to the Lorentzian lineshape difficult.
Since the interaction of the main 1605 cm$^{-1}$ peak with the
electronic continuum makes the lineshape broader and asymmetric it
was more accurately determined by a BWF fit. It is possible only
to resolve one peak using a BWF lineshape. We have fit the entire
pressure range using both fits (see Figure \ref{figure7}). At
lower pressures there is good agreement between the peak
frequencies determined by both the Lorentzian and the BWF fits,
however the values deviate at higher pressures. Our linear fits to
the frequency vs pressure data for the Lorentzian fits are shown
in Figure \ref{figure7} up to 40 kbar.

Since the PL emission from the aggregates and
the keto defects red shift, at higher pressures the 647.1 nm Raman
excitation line (1.92 eV) excites the tail of the emission (Figure
\ref{figure3}), which couples strongly to the phonons. We
therefore directly excite the emission without exciting the
backbone PL emission, excluding the possibility of this emission
being from excimers.

To determine the peak position, asymmetry parameter and linewidth
as a function of pressure, we fit the 1605 cm$^{-1}$ peak with a
BWF lineshape given by\cite{zhou}
\begin{equation}\label{1}
I(\omega)= I_{0} \: \frac{[(\omega - \omega_0)/q +
\Gamma]^2}{(\omega - \omega_0)^2 + \Gamma^2} \: ,
\end{equation}
where $\omega_{0}$ is the discrete phonon frequency, and $\Gamma$
is the width of the resonant interference between the continuum
and discrete scattering channels. The asymmetry parameter $(1/q)$
depends on the average electron-phonon matrix element $M$ and the
Raman matrix elements between the ground and excited states of the
phonon and electron.\cite{meera} The broadening parameter is given
by

\begin{equation}\label{2}
\Gamma=\pi{M^2}D(\omega),
\end{equation}
where $D(\omega)$ is the combined density of states for the
electronic transitions. For non-zero density of states it turns
out that when $q=-(\omega - \omega_0)/\Gamma$, the spectral
function $I(\omega)$ will reveal an "antiresonance" close to the
value of phonon frequency.\cite{klein}

\begin{figure}
\unitlength1cm
\begin{picture}(5.0,6.)
\put(-2.,-.8){ \epsfig{file=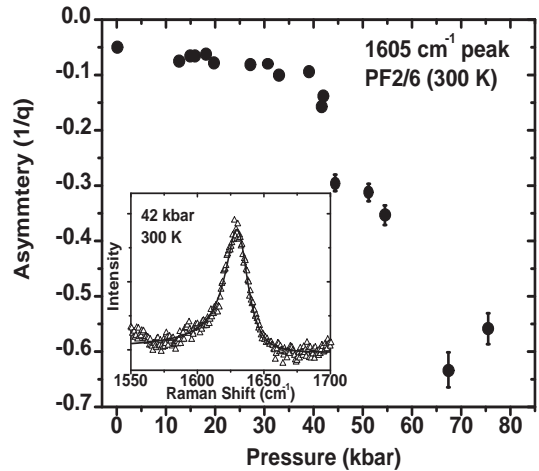, angle=0, width=8.2cm,
totalheight=7.5cm}}
\end{picture}
\caption{Asymmetry parameter($1/q$) of the 1605 cm$^{-1}$ Raman
peak versus pressure. $1/q$ is obtained by fits to the Raman peak
with a BWF line shape (Eq. {\ref{1}}). The inset shows a sample
BWF fit to the 1605 cm$^{-1}$ peak at 42 kbar.} \label{figure8}
\end{figure}

\begin{figure}
\unitlength1cm
\begin{picture}(5.0,7.0)
\put(-2.,-0.){ \epsfig{file=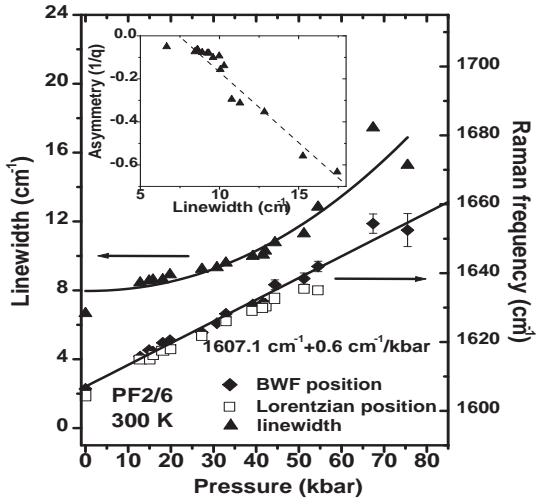, angle=0, width=7.8cm,
totalheight=7.5cm}}
\end{picture}
\caption{Linewidth and the peak position of the 1605 cm$^{-1}$
Raman peak obtained by fits to Eq.{\ref{1}}, as a function of
pressure. A linear fit to the BWF peak position of the 1605
cm$^{-1}$ Raman peak as a function of pressure yields a slope of
0.6 $\pm$ 0.02 cm$^{-1}$/kbar, as shown by the straight line. The
inset shows the asymmetry parameter to linearly vary with the
linewidth.} \label{figure9}
\end{figure}

Figure \ref{figure8} shows the asymmetry parameter versus pressure
for the 1605 cm$^{-1}$ Raman mode. The asymmetry is relatively
small up to about 35 kbar, with a {1/q} value between -0.005 and
0.1. Beyond 40 kbar however, the asymmetry increases rapidly and
{1/q} varies from -0.3 to -0.6. The inset shows a sample BWF fit
to Eq. {\ref{1}} (bold line) to the experimental data (open
triangles) for data at 42 kbar. The peak position and the
linewidths from the BWF fits as a function of pressure are plotted
in Figure \ref{figure9}. The linewidth shows a square law
dependence with pressure ($\Gamma \propto {P}^2$ ) (solid line in
Figure \ref{figure9}). The frequency positions vary almost
linearly with pressure. The inset of Figure \ref{figure9} shows
$1/q$ to linearly vary with the linewidth $\Gamma$.

Recently, optical techniques based on photoinduced infrared-active
vibrational (IRAV) modes have been used to study the
photogeneration and recombination dynamics of charged polarons in
conjugated polymers.\cite {miranda} \"{O}sterbacka  \textit{et
al}. have observed Fano-type antiresonances in the IRAV modes in a
series of $\pi$-conjugated polymers that are explained well by
extending the amplitude model beyond the adiabatic
limit.\cite{osterbacka} Our observation here is somewhat
different; the electronic continuum is from a combination of
aggregate and defect emissions that shift to lower energies with
increasing pressure. Figure 8 shows that q is negative, indicating
that the center frequency of the continuum lies below the discrete
mode frequency of ~1605 cm$^{-1}$ (0.2 eV). It is interesting to
point out that in $M_{3}C_{60}$ and superconducting cuprates,
$1/q$ value lies between -0.2 to -0.5,\cite{zhou} similar to our
results beyond 40 kbar. The high background and the interference
from the diamond Raman peak makes it difficult to analyze the
asymmetry of the other Raman peaks as a function of pressure. The
systematic appearance of the inverted 1530 cm$^{-1}$ peak at
higher pressures makes it a viable candidate for the IR-active
frequency that gets activated in the Raman spectrum due to a
lowering of the symmetry. Most probably the peak shows an S-like
behavior due to the antiresonance effect and therefore appears as
a negative peak.

Since the vibrational frequencies of a harmonic solid are
independent of compression, pressure induced changes in the Raman
spectrum provide insight into the anharmonicity of the solid state
potential.\cite{ferraro} In this light the linear shift of the
1605 cm$^{-1}$ peak as a function of pressure is not surprising.
The rate of shift is similar to other Raman-active frequencies in
PF2/6 which is an indication of the anharmonic potential. Although
one expects an additional shift in phonon frequency due to the BWF
interaction (the real part of the self energy) we cannot infer any
information about this parameter from our data. In materials where
the BWF interaction can be tuned, e.g. using uniaxial stress in
both n-type\cite{meera} and p-type Si, \cite{cardona} the real
part of the self energy due to interaction between the phonon and
the electronic continuum is determined by comparison with the pure
material. In this work, such a comparison is not possible owing to
a lack of a defect-free but identical material. Had this parameter
been large one would expect a significant deviation in the peak
position vs pressure behavior above 40 kbar where the BWF
character of the mode is pronounced.

However, the imaginary part of the self-energy, $\Gamma$, does
show a quadratic dependence with pressure. Furthermore, the
asymmetry parameter should be directly proportional to $\Gamma$ if
the average electron-phonon matrix element and the Raman matrix
elements between the ground and excited states are assumed to be
roughly constant.\cite{sathpathy} Indeed, this is what we find, as
shown in the inset of Figure \ref{figure9}. The negative slope
here is due to the negative q values that we obtain from the fit.
At present we are unclear of the physical significance of the
quadratic dependence of $\Gamma$ on pressure, which reflects on
the density of states of the continuum. A full-scale
band-structure calculation incorporating aggregate and defect
states should reveal the nature of the electronic density of
states.

\section{CONCLUSION} \label{sec:conclusion}

In conclusion, we have shown that hydrostatic pressure induces
very efficient energy transfer of the singlet excitons to the keto
defect sites mainly in bulk PF2/6. Enhanced intermolecular
interaction induces a strong aggregate-type emission in the film
sample, in addition to a keto emission. This may result from a
pinning of excitations in ordered regions in the film that
prevents excitation diffusion to the defect states. The Raman
modes harden with increasing pressures. The 1605 cm$^{-1}$ peak
exhibits an asymmetric lineshape, characteristic of a Breit-Wigner
Fano resonance, and some of the other peaks show an antiresonance
effect. This is indicative of a strong electron-phonon interaction
between the Raman phonons and the electronic continuum. We have
carried out a detailed BWF lineshape analysis for the 1605
cm$^{-1}$ peak as a function of pressure.

Both the PL and Raman studies from bulk PF2/6 show significant
changes at ~40 kbar. The aggregate and defect related emission
increases significantly and the asymmetry of the 1605 cm$^{-1}$
Raman peak becomes higher. This may be related to a phase
transition. Future work on different side-group substituted PFs
that have extraordinarily small keto defect concentration are in
progress to obtain insight into the morphological changes that
occur in PFs as a function of hydrostatic pressure.


\begin{acknowledgments}
S.G. acknowledges the donors of the American Chemical Society
Petroleum Research Fund ({\#}35735-GB5) for  partial support of
this research. U.S. thanks SONY International Europe, Stuttgart,
and the Deutsche Forschungsgemeinschaft (DFG) for financial
support. One of us (S.G.) thanks Michael Winokur for many valuable
insights and discussions.
\end{acknowledgments}



\end{document}